\documentclass[twocolumn,aps,prc,superscriptaddress,showpacs,floatfix]{revtex4-1}
\usepackage{url}
\usepackage{cancel}
\usepackage[colorlinks,linkcolor=blue,citecolor=blue,filecolor=black,urlcolor=blue]{hyperref}
\usepackage{epsfig,graphics}
\usepackage{graphicx}
\usepackage{dcolumn}
\usepackage{bm}
\usepackage[usenames]{color}
\usepackage{amssymb}
\usepackage{amsmath}
\usepackage{multirow}
\usepackage{float}
\usepackage{harpoon}
\usepackage{MnSymbol}
\usepackage{appendix}
\usepackage{color}
\usepackage{hyperref}
\usepackage{cleveref}
\usepackage{dutchcal}
\usepackage[normalem]{ulem}

\begin{document}

\title{Probing properties of nuclear spin-orbit interaction with nucleon spin polarization in intermediate-energy heavy-ion collisions}
\author{Jun Xu}\email{junxu@tongji.edu.cn}
\affiliation{School of Physics Science and Engineering, Tongji University, Shanghai 200092, China}
\begin{abstract}
The nucleon spin polarization perpendicular to the reaction plane ($P_y$) and along the beam direction ($P_z$) in Au+Au collisions at the beam energy of 100A MeV with different nuclear spin-orbit interactions has been studied based on a spin- and isospin-dependent Boltzmann-Uehling-Uhlenbeck (SIBUU) transport model. While the spin polarization is weaker with a weaker nuclear spin-orbit coupling as intuitively expected, a density-dependent nuclear spin-orbit coupling enhances the $P_y$ at large rapidities and leads to a less negative or large $P_y$ at high transverse momenta. The difference in the $P_y$ of free neutrons and protons at midrapidities and at small transverse momenta is sensitive to the isospin dependence of the nuclear spin-orbit interaction. While the $P_z$ is also affected by the properties of nuclear spin-orbit interaction in some sense, the behavior of the $P_y$ serves as a good probe of the strength, density dependence, and isospin dependence of nuclear spin-orbit interaction.
\end{abstract}
\maketitle


Spin polarization phenomena in high-energy heavy-ion collisions have attracted a considerable attention in the past few years. The global spin polarization and the spin alignment in non-central heavy-ion collisions perpendicular to the reaction plane are generated from the coupling of the particle spin with the total orbital angular momentum of the system. The global spin polarizations of $\Lambda$~\cite{STAR:2017ckg,PhysRevC.98.014910}, $\Omega$~\cite{PhysRevLett.126.162301}, and $\Xi$~\cite{PhysRevLett.126.162301} as well as the spin alignments of $\phi$ and $K^{\star 0}$~\cite{STAR:2022fan} have been measured through the angular distribution of their decays. The local spin polarization of $\Lambda$ along the beam direction has also been observed and shows certain azimuthal angular dependence~\cite{PhysRevLett.123.132301}, characterizing the rich physics of spin dynamics. The focus has now been transferred to the spin polarization at lower collision energies~\cite{PhysRevC.104.L061901,2022137506}, where the hadronic degree of freedom dominates the dynamics, and it has been observed that the spin polarization of $\Lambda$ is even stronger compared to that at higher collision energies.

In heavy-ion collisions dominated by nucleon degree of freedom, it is intuitively expected that the nuclear spin-orbit interaction is the main source of the particle spin polarization. It is well-known that the nuclear spin-orbit interaction is important in understanding the shell structure and the resulting magic number of finite nuclei~\cite{Mayer:1948zz,Mayer:1949pd,Haxel:1949fjd}. While the spin-orbit interaction is itself non-relativistic in, e.g., the Skyrme-Hartree-Fock (SHF) model, it is an essential component of the nuclear interaction, which can also be effectively obtained from the non-relativistic reduction of the Dirac equation in the relativistic mean-field (RMF) model. The strength, density dependence, and isospin dependence of the nuclear spin-orbit interaction remain still uncertain, and they are related to interesting topics in nuclear structures. From fitting the properties of light to heavy nuclei and taking into account the uncertainty due to the tensor force, the strength of the nuclear spin-orbit interaction ranges from 80 to 150 MeV fm$^5$~\cite{Lesinski:2007ys,Zalewski:2008is,Bender:2009ty}. The density and isospin dependence of the nuclear spin-orbit interaction are different in the SHF model compared to the reduction form in the RMF model. The density dependence is related to the bubble structure in finite nuclei~\cite{Sorlin_2013}. The kink of the charge radii of lead isotopes could be explained with proper density-dependent~\cite{PhysRevC.91.021302} or isospin-dependent ~\cite{Sharma:1994mim,Reinhard:1995zz} spin-orbit interaction. The discrepancy in extracting the symmetry energy from the neutron-skin thickness data for $^{208}$Pb and $^{48}$Ca measured by parity-violating electron scattering experiments could also be explained with a strong isospin dependence of the spin-orbit interaction~\cite{Yue:2024srj,Zhao:2024gjz}.

In previous studies, my collaborators and I have developed an SIBUU transport model, with the nucleon spin degree of freedom and the nuclear spin-orbit interaction incorporated in the mean-field approximation~\cite{Xu:2015kxa}. Due to the coupling of the nucleon spin with the orbital angular momentum in non-central heavy-ion collisions, nucleons with different spins perpendicular to the reaction plane are affected by different nuclear spin-orbit potentials, leading to their different collective flows~\cite{Xu:2012hh}. Particularly, the detailed behaviors of the flow splitting for spin-up and spin-down nucleons are related to the strength, density dependence, and isospin dependence of the nuclear spin-orbit interaction~\cite{Xia:2014qva}. The nucleon spin polarization can also be observed based on the same framework~\cite{Xia:2019whr}. It is thus of great interest to investigate the relation between properties of the nuclear spin-orbit interaction and detailed behaviors of the nucleon spin polarization in intermediate-energy heavy-ion collisions. The present study from the low-energy side could be helpful for understanding the experimental results on the $\Lambda$ spin polarization at higher collision energies.



Let's start from the spin-dependent BUU equation written as~\cite{Ring1980,Smith1989}
\begin{eqnarray}\label{BLE}
\frac{\partial \hat{f}}{\partial t}&+&\frac{i}{\hbar}\left [ \hat{\varepsilon},\hat{f}\right]+\frac{1}{2}\left ( \frac{\partial \hat{\varepsilon}}{\partial \vec{p}}\cdot \frac{\partial \hat{f}}{\partial \vec{r}}+\frac{\partial \hat{f}}{\partial \vec{r}}\cdot \frac{\partial \hat{\varepsilon}}{\partial \vec{p}}\right ) \nonumber\\
&-&\frac{1}{2}\left ( \frac{\partial \hat{\varepsilon}}{\partial
\vec{r}}\cdot \frac{\partial \hat{f}}{\partial
\vec{p}}+\frac{\partial \hat{f}}{\partial \vec{p}}\cdot
\frac{\partial \hat{\varepsilon}}{\partial \vec{r}}\right )=I_c,
\end{eqnarray}
where the single-particle energy $\hat{\varepsilon}$ and the phase-space distribution $\hat{f}$ are $2\times 2$ matrics, including both the spin-averaged and spin-dependent contributions, and $I_c$ represents the collision term. Based on a generalized test-particle method~\cite{Xia:2016xiw}, the left-hand side of the above equation leads to the equations of motion similar to the Hamiltonian equations, with the nucleon spin expectation vector as an additional degree of freedom.

The spin dynamics is dominated by the nuclear spin-orbit potential in the mean-field approach, and in the microscopic level it originates from the Skyrme-type spin-orbit nuclear interaction between nucleons at $\vec{r}_1$ and $\vec{r}_2$ expressed as~\cite{Vautherin:1971aw}
\begin{equation}\label{vsoi}
v_{so} = i W_0 (\vec{\sigma}_1+\vec{\sigma}_2) \cdot \vec{k}^\prime \times
\delta(\vec{r}_1-\vec{r}_2) \vec{k},
\end{equation}
where $W_0$ is the strength of the spin-orbit coupling with a default value of $150$ MeV fm$^5$ in the present study, $\vec{\sigma}_{1(2)}$ is the Pauli matrices, $\vec{k}=(\vec{p}_1-\vec{p}_2)/2$ is the relative momentum operator acting on the right with $\vec{p}=-i\nabla$, and $\vec{k}^\prime$ is the complex conjugate of $\vec{k}$ acting on the left. Based on the Hartree-Fock method~\cite{Engel:1975zz}, the contribution to the energy-density functional can be expressed as
\begin{eqnarray}\label{vso}
V_{so} &=& -\frac{W_0^\star}{2}[\alpha(\rho \nabla \cdot \vec{J} + \vec{s} \cdot \nabla \times \vec{j}) \notag\\
&+& \beta\sum_\tau (\rho_\tau \nabla \cdot \vec{J}_\tau + \vec{s}_\tau \cdot \nabla \times \vec{j}_\tau)],
\end{eqnarray}
where both time-even and time-odd contributions are included. In the above, $\tau=n,p$ is the isospin index, $\rho=\sum_\tau \rho_\tau$, $\vec{s}=\sum_\tau \vec{s}_\tau$, $\vec{j}=\sum_\tau \vec{j}_\tau$, and $\vec{J}=\sum_\tau \vec{J}_\tau$ are the number, spin, momentum, and spin-current densities, respectively. In the default case from Eq.~(\ref{vsoi}), $W_0^\star=W_0$, $\alpha=1$, and $\beta=1$ are obtained. To mimic the density dependence of the nuclear spin-orbit interaction~\cite{Xu:2012hh,Xia:2014qva,Xu:2015kxa}, instead of introducing a density-dependent term in the two-body spin-orbit interaction [Eq.~(\ref{vsoi})] as in, e.g., Ref.~\cite{PhysRevC.91.021302}, I set
\begin{equation}
W_0^\star = W_0 (\rho/\rho_0)^\gamma,
\end{equation}
with $\rho_0=0.16$ fm$^{-3}$ being the saturation density, and $\gamma=0$ and 1 representing different density dependencies. Results with $\alpha=2$ and $\beta=-1$ representing a different isospin dependency in Eq.~(\ref{vso}) will also be compared in the present study.

For the spin-independent part, the following potential energy density representing a momentum-independent potential is used
\begin{equation}
V_{MID} = \frac{a}{2}\left(\frac{\rho}{\rho_0}\right)+\frac{b}{c+1}\left(\frac{\rho}{\rho_0}\right)^{c+1} + E_{sym}^{pot}\left(\frac{\rho}{\rho_0}\right)^{\gamma_{sym}} \delta^2,
\end{equation}
with $\delta=(\rho_n-\rho_p)/\rho$ being the isospin asymmetry, and the parameters $a=-209.2$ MeV, $b=156.4$ MeV, $c=1.35$, $E_{sym}^{pot}=18$ MeV, and $\gamma_{sym}=2/3$ reproducing empirical nuclear matter properties.

A lattice Hamiltonian method~\cite{Lenk:1989zz} is used to calculate the average number, spin, momentum, and spin-current densities from the nucleon phase-space distribution~\cite{Xia:2019whr}, i.e.,
\begin{eqnarray}
\rho(\vec{r}_{\alpha})&=&\sum_{i}S(\vec{r}_{\alpha}-\vec{r}_i),\\
\vec{s}(\vec{r}_{\alpha})&=&\sum_{i}\vec{\sigma}_iS(\vec{r}_{\alpha}-\vec{r}_i),\\
\vec{j}(\vec{r}_{\alpha})&=&\sum_{i}\vec{p}_iS(\vec{r}_{\alpha}-\vec{r}_i),\\
\vec{J}(\vec{r}_{\alpha})&=&\sum_{i}\left(\vec{p}_i \times \vec{\sigma}_i\right)S(\vec{r}_{\alpha}-\vec{r}_i),
\end{eqnarray}
where $\vec{r}_i$, $\vec{p}_i$, and $\vec{\sigma}_i$ are respectively the position, momentum, and spin expectation vector of the $i$th nucleon, $\vec{r}_{\alpha}$ is the position of the three-dimensional lattice site $\alpha$, and $S$ is the shape function describing the contribution of a test particle at $\vec{r}_i$ to the average density at $\vec{r}_{\alpha}$, i.e.,
\begin{eqnarray}
S(\vec{r})=\frac{1}{N(nl)^6}g(x)g(y)g(z)
\end{eqnarray}
with
\begin{eqnarray}
g(q)=(nl-|q|)\Theta(nl-|q|).
\end{eqnarray}
In the above, $N$ is the number of parallel events, i.e., the test-particle number per nucleon, $l$ is the lattice spacing, $n$ determines the range of $S$, and $\Theta$ is the Heaviside function. I adopt $N=400$, $l=1$ fm, and $n=2$ in the present study. The Hamiltonian of the whole system is
\begin{equation}
H=\sum_{i}\frac{\vec{p}_{i}^{2}}{2m}+Nl^3\sum_\alpha [V_{MID}(\vec{r}_{\alpha}) + V_{so}(\vec{r}_{\alpha})].
\end{equation}
The canonical equations of motion in the lattice Hamiltonian framework become
\begin{eqnarray}
\frac{d\vec{r}_i}{dt} &=& \frac{\partial H}{\partial \vec{p}_i}, \label{dr}\\
\frac{d\vec{p}_i}{dt} &=& - \frac{\partial H}{\partial \vec{r}_i}, \label{dp}\\
\frac{d\vec{\sigma}_i}{dt} &=& \frac{1}{i} [\vec{\sigma}_i, H]. \label{proc}
\end{eqnarray}
It is seen that nucleons with different spins are affected by the different spin-orbit potential according to Eqs.~(\ref{dr}) and (\ref{dp}). Equation~(\ref{proc}) can be understood by first assuming that both $\sigma_i$ and $H$ are operators and then replacing them with their semiclassical correspondence quantities after the commutation calculation, and it actually leads to
\begin{equation}
\frac{d\vec{\sigma}_i}{dt} = 2 \vec{h} \times \vec{\sigma}_i,
\end{equation}
with
\begin{equation}
\vec{h} = -\frac{W_{0}^\star}{2}\nabla \times (\alpha\vec{j}+\beta\vec{j}_{\tau}) +\frac{W_{0}^\star}{2}[\nabla(\alpha\rho+\beta\rho_{\tau}) \times \vec{p}].
\end{equation}
Equation~(\ref{proc}) describes the precession of the spin expectation direction. As pointed out in the previous study~\cite{Xu:2012hh}, the time-odd terms ($\vec{j}$ and $\vec{s}$ terms in Eq.~(\ref{vso})) are generally stronger than the time-even terms ($\rho$ and $\vec{J}$ terms in Eq.~(\ref{vso})), and dominate the dynamics in most cases.

For the collision term ($I_c$ in Eq.~(\ref{BLE})), different spin-singlet and spin-triplet neutron-neutron (proton-proton) and neutron-proton elastic scattering cross sections as parametrized in Ref.~\cite{Xia:2017dbx} are used, and they were extracted from the phase-shift analyses of nucleon-nucleon scatterings in free space~\cite{PhysRevC.15.1002}. The spin- and isospin-dependent Pauli blocking has also been implemented with more specific local phase-space cells for nucleons with different spin and isospin states~\cite{PhysRevC.109.014615}. The spins of nucleons after a successful collisions are assumed to be unchanged, from the estimate of the spin change in the presence of the Argonne potential~\cite{Wiringa:1994wb,PhysRevC.109.014615}.


The present study focuses on mid-central ($\text{b}=8$ fm) and mid-peripheral ($\text{b}=12$ fm) Au+Au collisions at the beam energy of 100A MeV. The simulation stops at $t=100$ fm/c when the two colliding nuclei have stopped interacting with each other. The spin polarization is analyzed based on the information of final-state free nucleons, around which the local density is below $\rho_0/8$.


\begin{figure}[ht]
\includegraphics[width=1\linewidth]{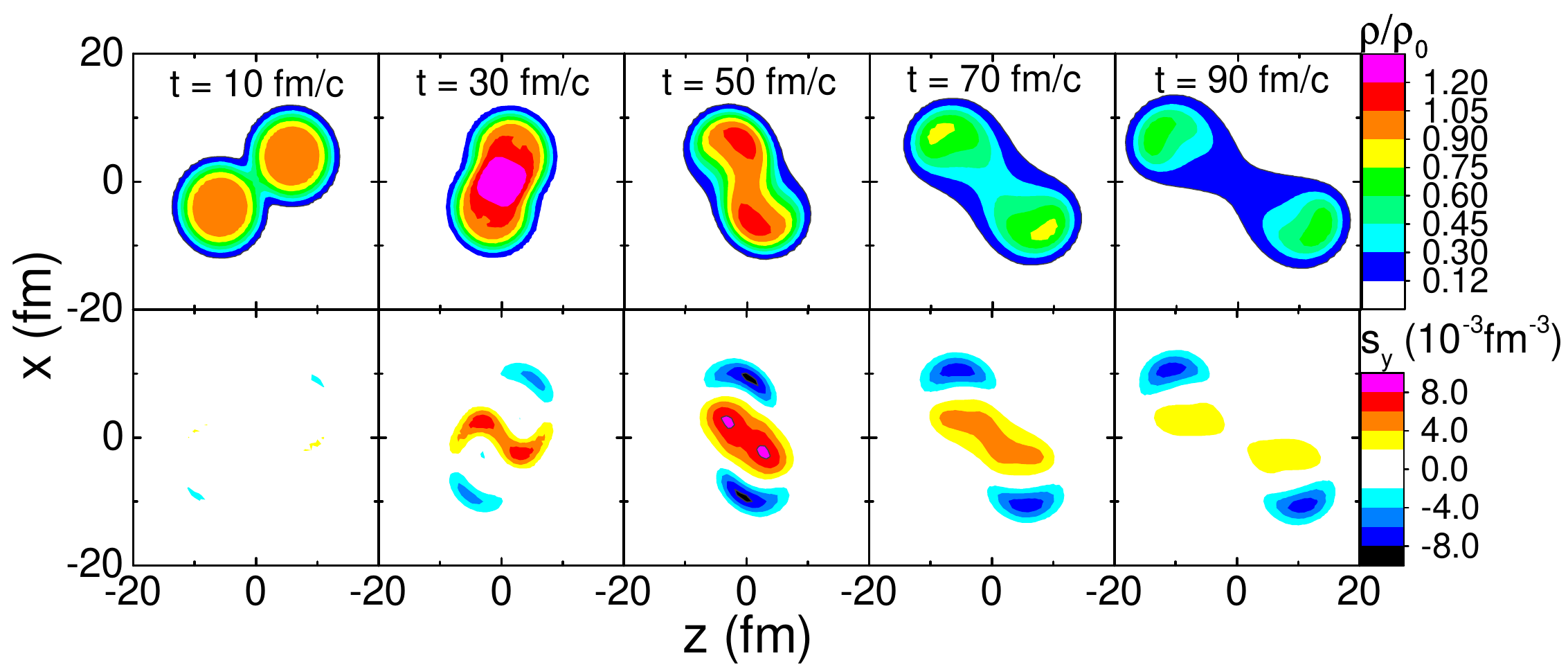}\\
\includegraphics[width=1\linewidth]{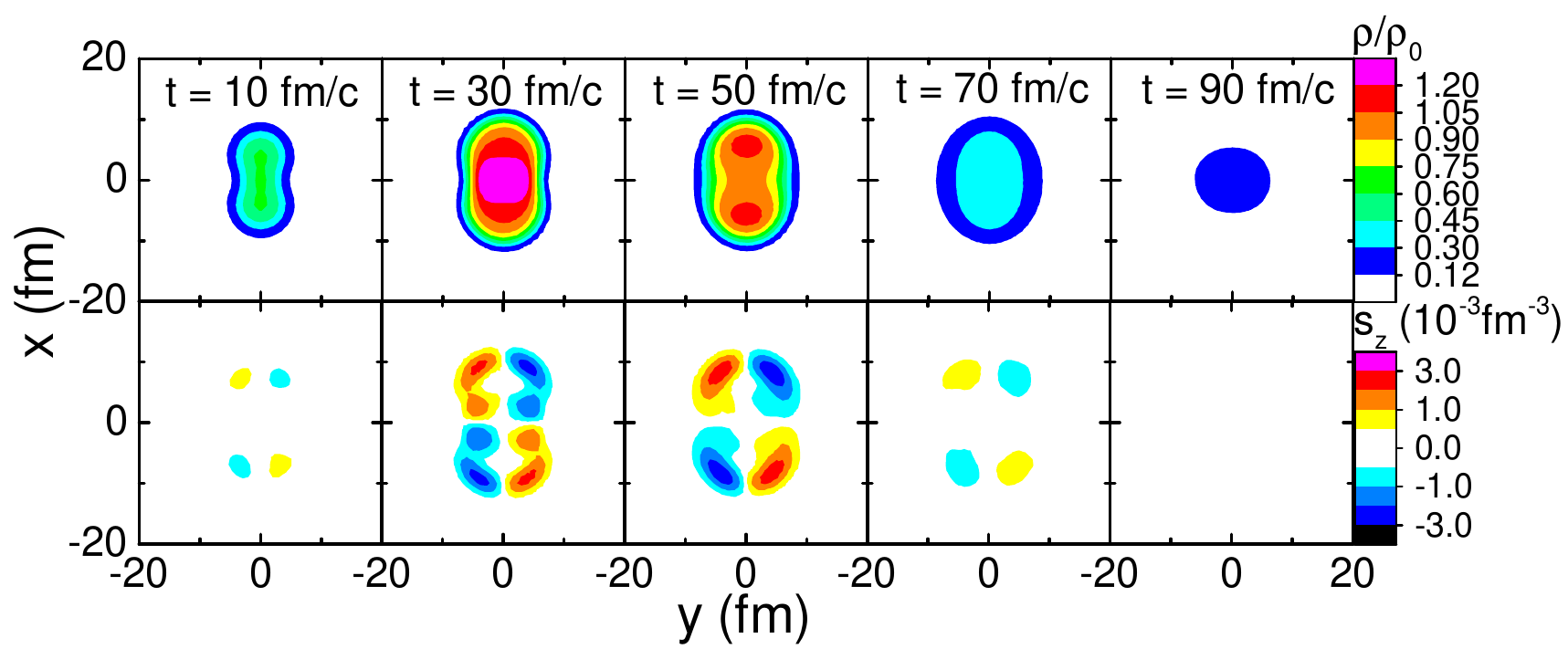}
\caption{\label{den} Upper: Contours of the reduced density $\rho/\rho_0$ (first row) and the $y$ component of the spin density in the x-o-z plane in mid-central collisions from $W_0^\star=150$ MeV fm$^5$. Lower: Contours of the reduced density $\rho/\rho_0$ (first row) and the $z$ component of the spin density in the x-o-y plane in mid-central collisions from $W_0^\star=150$ MeV fm$^5$.}
\end{figure}

Before showing the spin polarization, we first display the evolutions of the nucleon number density and spin density in mid-central collisions in Fig.~\ref{den}, where the upper panels show the density evolution in the reaction plane, and the lower panels show the density evolution in the transverse plane. Before the projectile and target interact with each other, no spin polarization is generated due to the cancellation of the time-odd term $-(\nabla\times\vec{j})_y$ and the time-even term $(\vec{p}\times\nabla\rho)_y$ in the spin-orbit potential, preserving the Galilean invariance. Later on, while the net spin-orbit potential for spin $\pm y$ nucleons still vanishes in the spectator region, the time-odd potential overwhelms the time-even potential in the participant region, leading to a more attractive (repulsive) potential for spin $+y$ ($-y$) nucleons. This attracts the spin $+y$ nucleons from the spectator region, and repels the spin $-y$ nucleons to the spectator region, generating a positive $y$-component of the spin density $s_y$ in the participant matter but a negative $s_y$ in the spectator region~\cite{Xia:2019whr}. On the other hand, the $z$-component of the spin density $s_z$ in both the participant and spectator regions show some azimuthal angular dependence, dominated respectively by the time-odd term and the time-even term~\cite{Xia:2019whr}.

\begin{figure}[ht]
\includegraphics[width=0.9\linewidth]{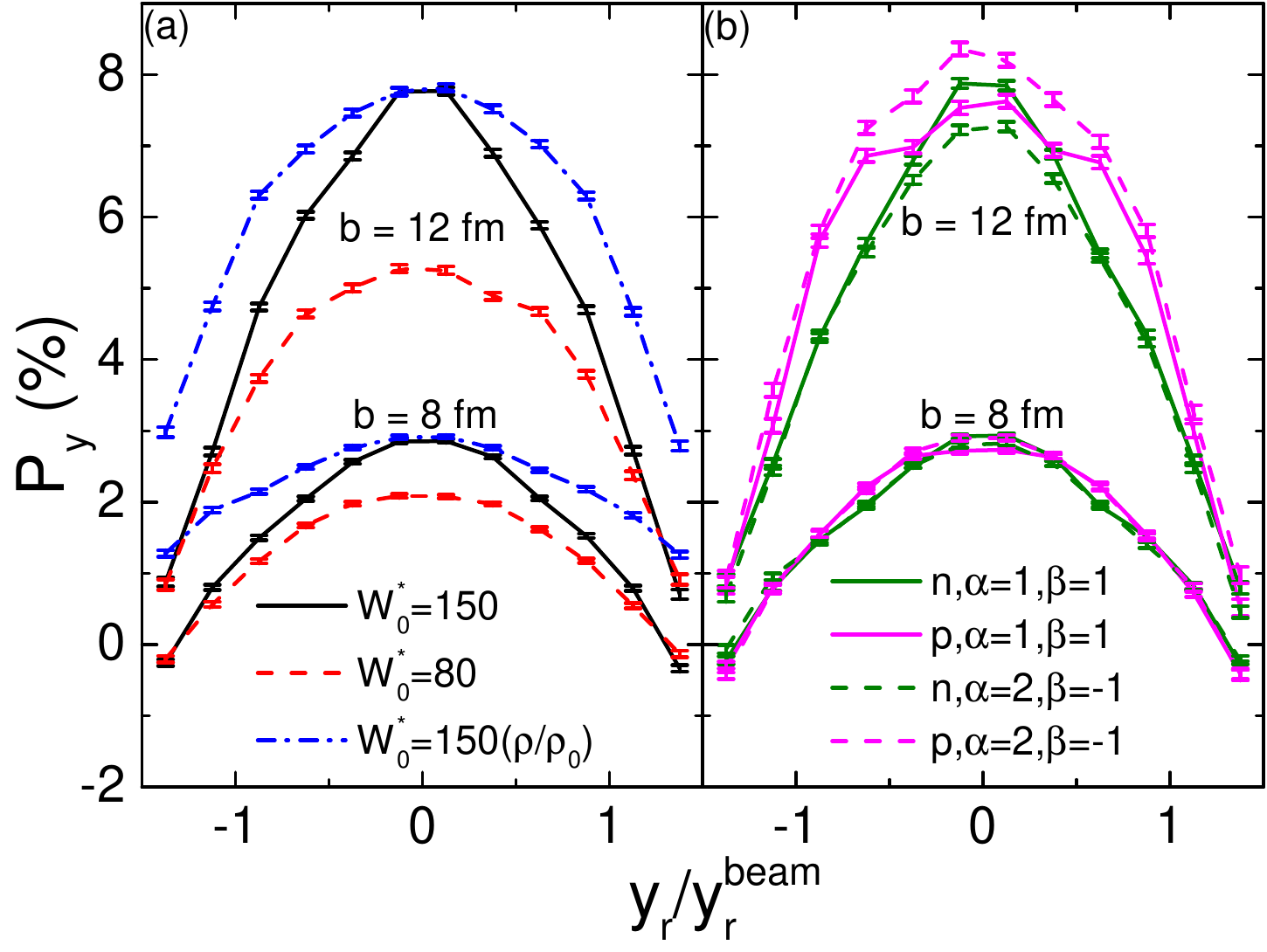}
\caption{\label{py-rap} Rapidity dependence of $P_y$ of free nucleons from different spin-orbit coupling coefficients (a) and of free neutrons and protons from different isospin dependencies of the spin-orbit interaction (b) in mid-central and mid-peripheral collisions.}
\end{figure}

From different nuclear spin-orbit interactions, the spin polarizations $P_y$ in $y$ direction as functions of the reduced rapidity $y_\text{r}/y_\text{r}^{\text{beam}}$, with $y_\text{r}^{\text{beam}}$ being the beam rapidity in the center-of-mass frame of Au+Au collisions, are compared in Fig.~\ref{py-rap}. Here the $P_y$ is defined as
\begin{equation}\label{py}
P_y = \frac{N_{\mathcal{s}_y=+\frac{1}{2}}-N_{\mathcal{s}_y=-\frac{1}{2}}}{N_{\mathcal{s}_y=+\frac{1}{2}}+N_{\mathcal{s}_y=-\frac{1}{2}}},
\end{equation}
where $N_{\mathcal{s}_y=\pm\frac{1}{2}}$ is the number of free nucleons at spin state $\mathcal{s}_y = \pm\frac{1}{2}$. In Fig.~\ref{py-rap}(a), it is seen that the $P_y$ is generally larger at midrapidities dominated by the spin density in the participant region, and smaller at large rapidities affected by the spin density in the spectator region. Since the spin-orbit potential depends on the density gradient, it is stronger in mid-peripheral collisions than in mid-central collisions, leading to a stronger spin polarization in the former case. As expected, a stronger spin-orbit coupling coefficient $W_0^\star=150$ MeV fm$^5$ leads to a larger $P_y$ compared to $W_0^\star=80$ MeV fm$^5$, while it seems that the growth of $P_y$ with $W_0^\star$ is slower than a linear trend. It is interesting to see that a density-dependent spin-orbit coupling coefficient $W_0^\star=150(\rho/\rho_0)$ MeV fm$^5$ leads to a similar $P_y$ at midrapidities but a larger $P_y$ at large rapidities, compared to the case of $W_0^\star=150$ MeV fm$^5$. Note that a considerable amount of free nucleons at large rapidities emit from the participant matter away from the central region but close to the spectators. Around the regions between the participant matter and the spectators, as shown in the upper panel of Fig.~\ref{den} at $t=30$ fm/c, the density could be slightly higher than $\rho_0$, so the spin-orbit potential determined by the density gradient there can be even stronger for $W_0^\star=150(\rho/\rho_0)$ MeV fm$^5$ than for $W_0^\star=150$ MeV fm$^5$. Thus, the rapidity dependence of the $P_y$ could be a probe of the density dependence of the nuclear spin-orbit interaction. The $P_y$ of free neutrons and protons from different isospin dependencies are compared in Fig.~\ref{py-rap}(b). At midrapidities where nucleons are emitted from the neutron-rich participant region, it is seen that neutrons have a larger $P_y$ than protons for $\alpha=1$ and $\beta=1$, but the inverse is observed for $\alpha=2$ and $\beta=-1$. The effect is larger in peripheral collisions, where the neutron skin of colliding nuclei may also play a role. Thus, the difference in the $P_y$ of neutrons and protons at midrapidities could be a probe of the isospin dependence of the nuclear spin-orbit interaction.

\begin{figure}[ht]
\includegraphics[width=0.9\linewidth]{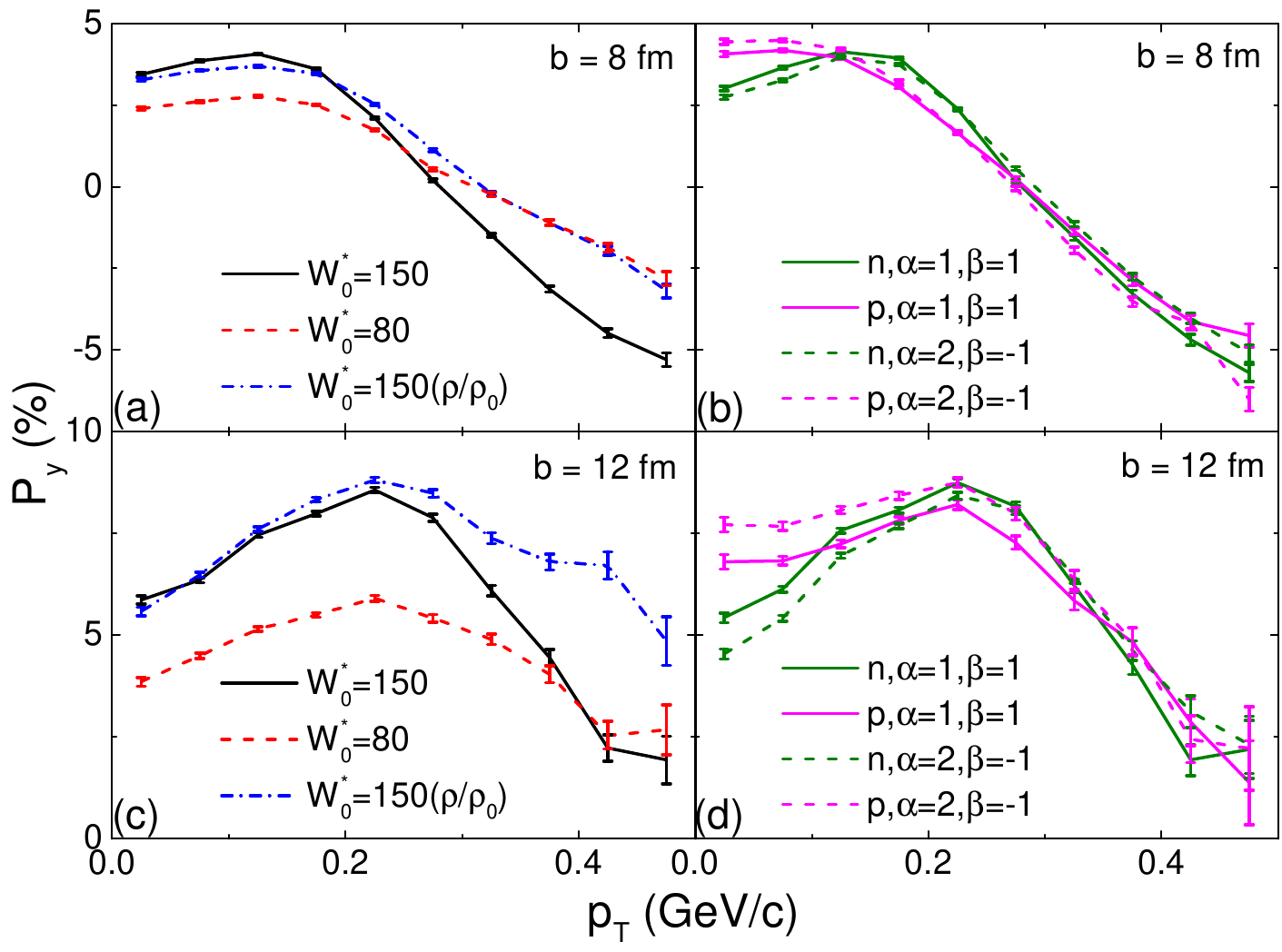}
\caption{\label{py-pt} Transverse momentum dependence of $P_y$ of free nucleons from different spin-orbit coupling coefficients [(a), (c)] and of free neutrons and protons from different isospin dependencies of the spin-orbit interaction [(b), (d)] in mid-central (upper) and mid-peripheral (lower) collisions. Only midrapidity ($|y_\text{r}|<y_\text{r}^{\text{beam}}/2$) nucleons are considered in the analysis.}
\end{figure}

Figure~\ref{py-pt} compares the transverse momentum ($p_T$) dependence of $P_y$ of free nucleons at midrapidities for the same scenarios as in Fig.~\ref{py-rap}. High-$p_T$ nucleons emit in the early stage from the participant region. In mid-central collisions, these nucleons are blocked by the spectator matter which has a negative $s_y$, as shown in the upper panel of Fig.~\ref{den}, and they eventually have a negative $P_y$. In mid-peripheral collisions, due to the overall stronger $P_y$ in the participant matter and weaker blocking effect from the spectator matter, high-$p_T$ nucleons eventually have a smaller but still positive $P_y$, compared to the situation in mid-central collisions. As expected, $W_0^\star=80$ MeV fm$^5$ leads to an overall weaker spin polarization, compared to $W_0^\star=150$ MeV fm$^5$. $W_0^\star=150(\rho/\rho_0)$ MeV fm$^5$ leads to a less negative $P_y$ at high $p_T$ in mid-central collisions, and enhances the $P_y$ at high $p_T$ in mid-peripheral collisions, compared to $W_0^\star=150$ MeV fm$^5$. This is due to the stronger spin-orbit potential for $W_0^\star=150(\rho/\rho_0)$ MeV fm$^5$ in the high-density participant matter, which leads to stronger $P_y$ for high-$p_T$ nucleons before they are affected by the negative $s_y$ of the spectator matter. At small $p_T$ where nucleons are emitted from the more neutron-rich low-density region, the effect of the isospin dependence of the nuclear spin-orbit interaction is expected to be stronger, while the detailed spin dynamics also depends on the density gradient. It is interesting to see that free protons have a larger $P_y$ than free neutrons, and such splitting is even larger for $\alpha=2$ and $\beta=-1$. At high $p_T$, the effect of the isospin dependence of the nuclear spin-orbit interaction on the difference in $P_y$ of neutrons and protons is not so obvious.

\begin{figure}[ht]
\includegraphics[width=0.9\linewidth]{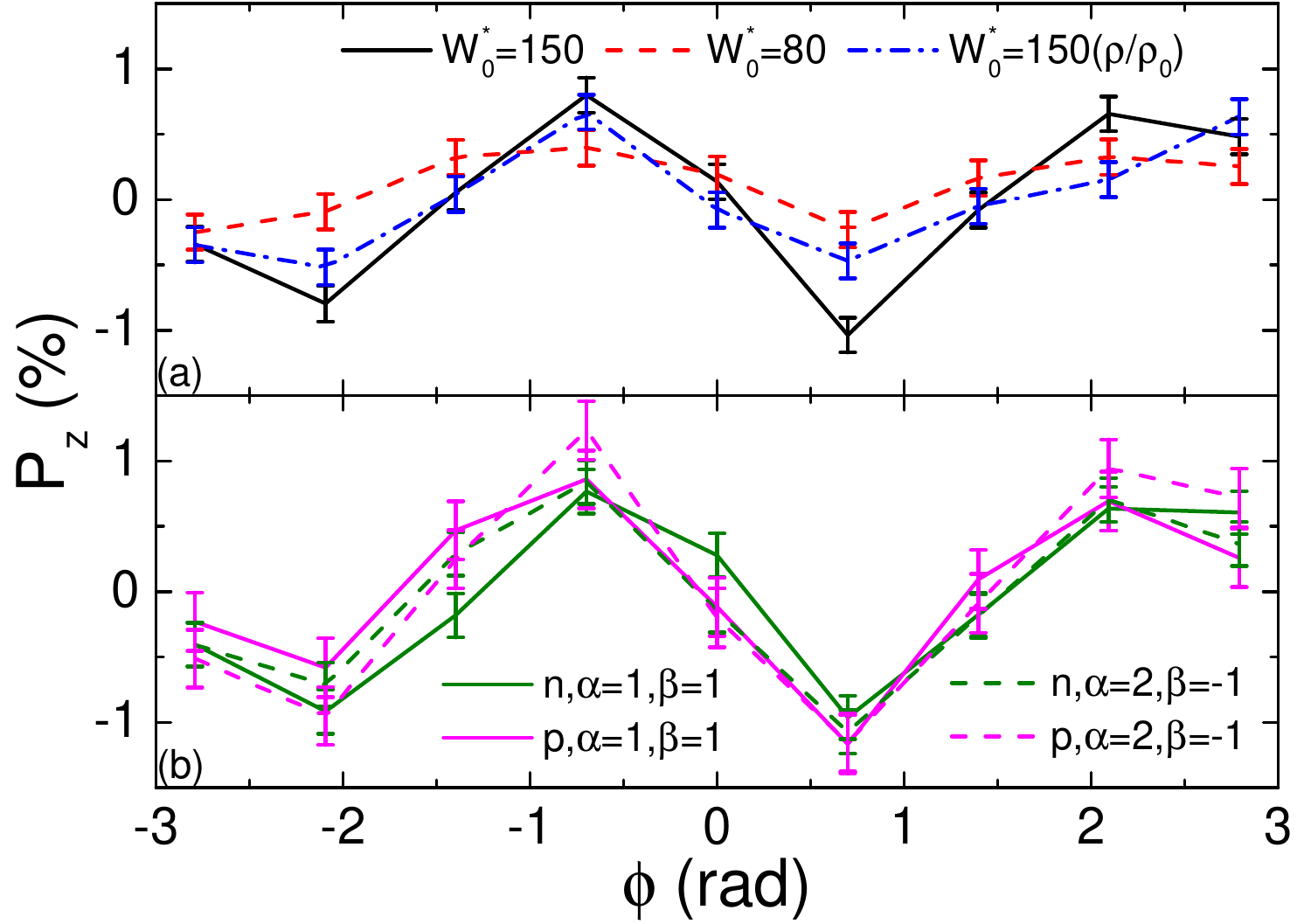}
\caption{\label{pz-phi} Azimuthal angular dependence of $P_z$ of free nucleons from different spin-orbit coupling coefficients (a) and of free neutrons and protons from different isospin dependencies of the spin-orbit interaction (b) in mid-central collisions. Only midrapidity ($|y_\text{r}|<y_\text{r}^{\text{beam}}/2$) energetic ($p_T>0.3$ GeV/c) nucleons are considered in the analysis.}
\end{figure}

The local spin polarization of $\Lambda$ hyperons in the longitudinal direction, especially its ``sign'' problem in the azimuthal angular dependence, has attracted considerable attention~\cite{PhysRevLett.123.132301,BECATTINI2021136519,PhysRevLett.120.012302,PhysRevLett.127.142301,PhysRevLett.125.062301,PhysRevResearch.1.033058}. For the illustration purpose, Fig.~\ref{pz-phi} displays the azimuthal angular dependence of the longitudinal spin polarization $P_z$ of free energetic nucleons at midrapidities in mid-central collisions. Here the azimuthal angle is calculated from $\phi = \text{atan2}(p_y,p_x)$, with $p_y$ and $p_x$ being respectively the nucleon momentum in $y$ and $x$ direction, and $P_z$ is defined similarly to $P_y$ as in Eq.~(\ref{py}). It is seen that $P_z$ has negative peaks at $\phi=-3\pi/4$ and $\pi/4$ and positive peaks at $\phi=-\pi/4$ and $3\pi/4$, consistent with the distribution of $s_z$ shown in the lower panel of Fig.~\ref{den}. A smaller $W_0^\star$ or a density-dependent one generally weakens the $P_z$, as shown in Fig.~\ref{pz-phi}(a). In addition, the $P_z$ of free neutrons and protons show similar behavior within statistical error even for different isospin dependencies of the nuclear spin-orbit interaction.

\begin{figure}[ht]
\includegraphics[width=0.9\linewidth]{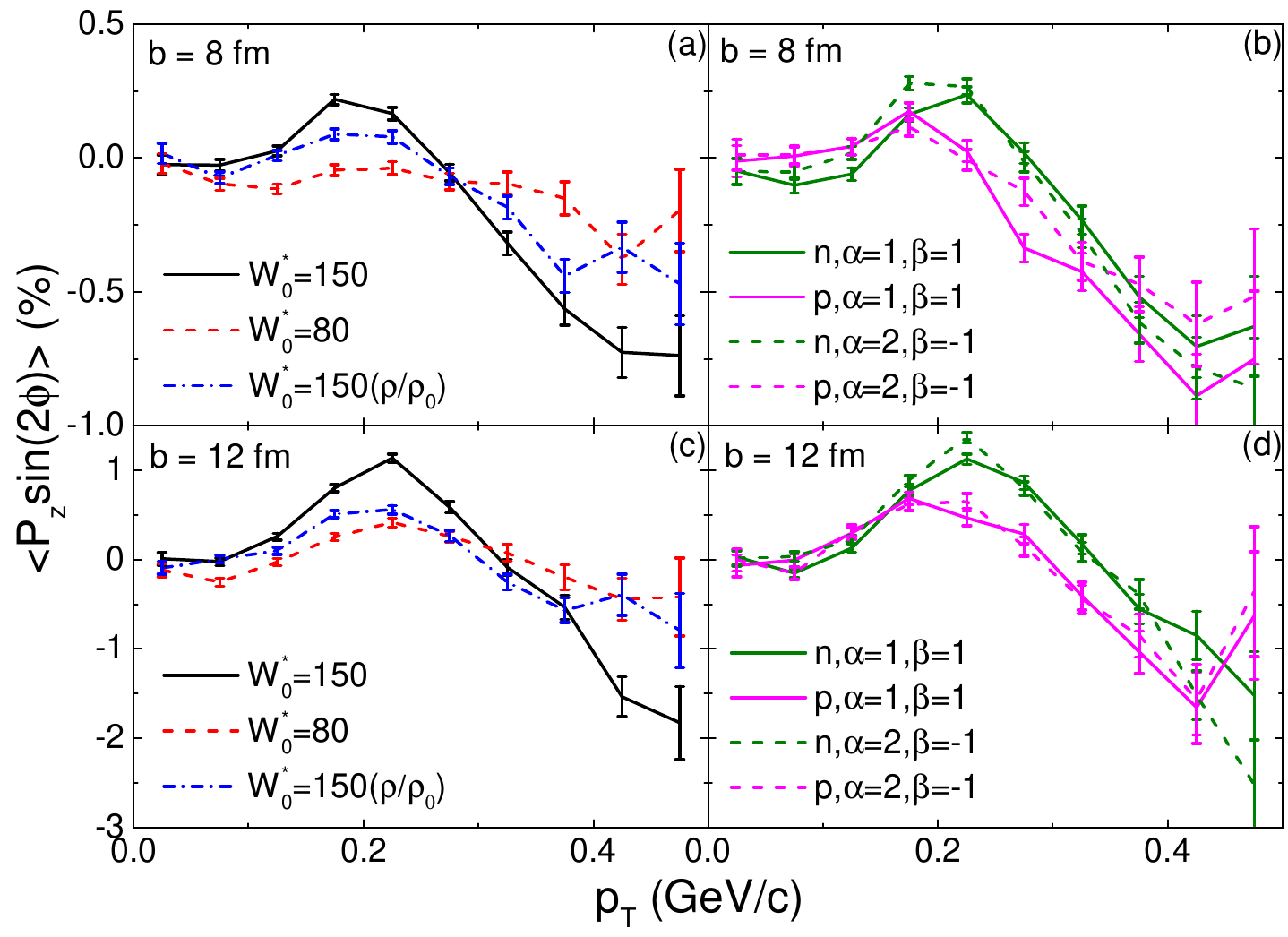}
\caption{\label{pz-phi-pt} Transverse momentum dependence of the second Fourier sine coefficient of the longitudinal spin polarization of free nucleons from different spin-orbit coupling coefficients [(a), (c)] and of free neutrons and protons from different isospin dependencies of the spin-orbit interaction [(b), (d)] in mid-central (upper) and mid-peripheral (lower) collisions. Only midrapidity ($|y_\text{r}|<y_\text{r}^{\text{beam}}/2$) nucleons are considered in the analysis.}
\end{figure}

The magnitude of the local longitudinal spin polarization can be characterized by its second Fourier sine coefficient $\langle P_z \sin(2\phi) \rangle$, whose transverse momentum dependencies for different scenarios are compared in Fig.~\ref{pz-phi-pt}. Similar to the situation of $P_y$, $\langle P_z \sin(2\phi) \rangle$ is stronger in mid-peripheral collisions than in mid-central collisions. Nucleons at high $p_T$ emit in the early stage of the collision, and thus show a similar azimuthal angular distribution as that of $s_z$ at $t=30-50$ fm/c shown in the lower panel of Fig.~\ref{den}. Nucleons at low and intermediate $p_T$ emit in the late stage of the collision from the neck region, and have a small and positive $\langle P_z \sin(2\phi) \rangle$, similar to that of $\Lambda$ hyperons observed in relativistic heavy-ion collisions~\cite{PhysRevLett.123.132301}. Thus the ``sign'' in the azimuthal angular dependence of $P_z$ depends on both the transverse momentum and the centrality of the collision. As already seen in Fig.~\ref{pz-phi}, a smaller $W_0^\star$ or a density-dependent one generally leads to a weaker local spin polarization and thus a smaller $\langle P_z \sin(2\phi) \rangle$. While free neutrons generally has a larger $\langle P_z \sin(2\phi) \rangle$ than free protons at intermediate $p_T$, it is difficult to probe the isospin dependence of the nuclear spin-orbit interaction by comparing the $P_z$ of neutrons and protons.


To summarize, the spin polarizations of free nucleons perpendicular to the reaction plane ($P_y$) and along the beam direction ($P_z$) in Au+Au collisions at the beam energy of 100A MeV have been studied from different nuclear spin-orbit interactions based on the SIBUU transport model. Generally, a stronger spin polarization is observed in mid-peripheral collisions than in mid-central collisions, and both $P_y$ and $P_z$ are weaker with a weaker nuclear spin-orbit coupling. A density-dependent nuclear spin-orbit coupling enhances the $P_y$ at large rapidities and leads to a less negative or larger $P_y$ at high transverse momenta. The difference in the $P_y$ of neutrons and protons at midrapidities and at small transverse momenta is found to be sensitive to the isospin dependence of the nuclear spin-orbit interaction. The ``sign'' in the azimuthal angular dependence of the $P_z$ depends on both the transverse momentum and the centrality of the collision. The $P_z$ is generally weaker with a density-dependent nuclear spin-orbit coupling, and it is difficult to probe the isospin dependence of the nuclear spin-orbit interaction by comparing the $P_z$ of neutrons and protons. Overall, the $P_y$ of free nucleons can provide more information of the nuclear spin-orbit interaction compared to the $P_z$.

This work is supported by the National Key Research and Development Program of China under Grant No. 2022YFA1602404, the Strategic Priority Research Program of the Chinese Academy of Sciences under Grant No. XDB34030000, the National Natural Science Foundation of China under Grant Nos. 12375125, 11922514, and 11475243, and the Fundamental Research Funds for the Central Universities.

\bibliography{sibuu}

\begin{thebibliography}{38}%
\makeatletter
\providecommand \@ifxundefined [1]{%
 \@ifx{#1\undefined}
}%
\providecommand \@ifnum [1]{%
 \ifnum #1\expandafter \@firstoftwo
 \else \expandafter \@secondoftwo
 \fi
}%
\providecommand \@ifx [1]{%
 \ifx #1\expandafter \@firstoftwo
 \else \expandafter \@secondoftwo
 \fi
}%
\providecommand \natexlab [1]{#1}%
\providecommand \enquote  [1]{``#1''}%
\providecommand \bibnamefont  [1]{#1}%
\providecommand \bibfnamefont [1]{#1}%
\providecommand \citenamefont [1]{#1}%
\providecommand \href@noop [0]{\@secondoftwo}%
\providecommand \href [0]{\begingroup \@sanitize@url \@href}%
\providecommand \@href[1]{\@@startlink{#1}\@@href}%
\providecommand \@@href[1]{\endgroup#1\@@endlink}%
\providecommand \@sanitize@url [0]{\catcode `\\12\catcode `\$12\catcode
  `\&12\catcode `\#12\catcode `\^12\catcode `\_12\catcode `\%12\relax}%
\providecommand \@@startlink[1]{}%
\providecommand \@@endlink[0]{}%
\providecommand \url  [0]{\begingroup\@sanitize@url \@url }%
\providecommand \@url [1]{\endgroup\@href {#1}{\urlprefix }}%
\providecommand \urlprefix  [0]{URL }%
\providecommand \Eprint [0]{\href }%
\providecommand \doibase [0]{http://dx.doi.org/}%
\providecommand \selectlanguage [0]{\@gobble}%
\providecommand \bibinfo  [0]{\@secondoftwo}%
\providecommand \bibfield  [0]{\@secondoftwo}%
\providecommand \translation [1]{[#1]}%
\providecommand \BibitemOpen [0]{}%
\providecommand \bibitemStop [0]{}%
\providecommand \bibitemNoStop [0]{.\EOS\space}%
\providecommand \EOS [0]{\spacefactor3000\relax}%
\providecommand \BibitemShut  [1]{\csname bibitem#1\endcsname}%
\let\auto@bib@innerbib\@empty
\bibitem [{\citenamefont {Adamczyk}\ \emph {et~al.}(2017)\citenamefont
  {Adamczyk} \emph {et~al.}}]{STAR:2017ckg}%
  \BibitemOpen
  \bibfield  {author} {\bibinfo {author} {\bibfnamefont {L.}~\bibnamefont
  {Adamczyk}} \emph {et~al.} (\bibinfo {collaboration} {STAR}),\ }\href
  {\doibase 10.1038/nature23004} {\bibfield  {journal} {\bibinfo  {journal}
  {Nature}\ }\textbf {\bibinfo {volume} {548}},\ \bibinfo {pages} {62}
  (\bibinfo {year} {2017})}\BibitemShut {NoStop}%
\bibitem [{\citenamefont {Adam~{\it et al.}}(2018)}]{PhysRevC.98.014910}%
  \BibitemOpen
  \bibfield  {author} {\bibinfo {author} {\bibfnamefont {J.}~\bibnamefont
  {Adam~{\it et al.}}} (\bibinfo {collaboration} {STAR Collaboration}),\ }\href
  {\doibase 10.1103/PhysRevC.98.014910} {\bibfield  {journal} {\bibinfo
  {journal} {Phys. Rev. C}\ }\textbf {\bibinfo {volume} {98}},\ \bibinfo
  {pages} {014910} (\bibinfo {year} {2018})}\BibitemShut {NoStop}%
\bibitem [{\citenamefont {Adam~{\it et al.}}(2021)}]{PhysRevLett.126.162301}%
  \BibitemOpen
  \bibfield  {author} {\bibinfo {author} {\bibfnamefont {J.}~\bibnamefont
  {Adam~{\it et al.}}} (\bibinfo {collaboration} {STAR Collaboration}),\ }\href
  {\doibase 10.1103/PhysRevLett.126.162301} {\bibfield  {journal} {\bibinfo
  {journal} {Phys. Rev. Lett.}\ }\textbf {\bibinfo {volume} {126}},\ \bibinfo
  {pages} {162301} (\bibinfo {year} {2021})}\BibitemShut {NoStop}%
\bibitem [{\citenamefont {Abdallah}\ \emph {et~al.}(2023)\citenamefont
  {Abdallah} \emph {et~al.}}]{STAR:2022fan}%
  \BibitemOpen
  \bibfield  {author} {\bibinfo {author} {\bibfnamefont {M.~S.}\ \bibnamefont
  {Abdallah}} \emph {et~al.} (\bibinfo {collaboration} {STAR}),\ }\href
  {\doibase 10.1038/s41586-022-05557-5} {\bibfield  {journal} {\bibinfo
  {journal} {Nature}\ }\textbf {\bibinfo {volume} {614}},\ \bibinfo {pages}
  {244} (\bibinfo {year} {2023})}\BibitemShut {NoStop}%
\bibitem [{\citenamefont {Adam~{\it et al.}}(2019)}]{PhysRevLett.123.132301}%
  \BibitemOpen
  \bibfield  {author} {\bibinfo {author} {\bibfnamefont {J.}~\bibnamefont
  {Adam~{\it et al.}}} (\bibinfo {collaboration} {STAR Collaboration}),\ }\href
  {\doibase 10.1103/PhysRevLett.123.132301} {\bibfield  {journal} {\bibinfo
  {journal} {Phys. Rev. Lett.}\ }\textbf {\bibinfo {volume} {123}},\ \bibinfo
  {pages} {132301} (\bibinfo {year} {2019})}\BibitemShut {NoStop}%
\bibitem [{\citenamefont {Abdallah~{\it et al.}}(2021)}]{PhysRevC.104.L061901}%
  \BibitemOpen
  \bibfield  {author} {\bibinfo {author} {\bibfnamefont {M.~S.}\ \bibnamefont
  {Abdallah~{\it et al.}}} (\bibinfo {collaboration} {STAR Collaboration}),\
  }\href {\doibase 10.1103/PhysRevC.104.L061901} {\bibfield  {journal}
  {\bibinfo  {journal} {Phys. Rev. C}\ }\textbf {\bibinfo {volume} {104}},\
  \bibinfo {pages} {L061901} (\bibinfo {year} {2021})}\BibitemShut {NoStop}%
\bibitem [{\citenamefont {Abou Yassine~{\it et al.}}(2022)}]{2022137506}%
  \BibitemOpen
  \bibfield  {author} {\bibinfo {author} {\bibfnamefont {R.}~\bibnamefont {Abou
  Yassine~{\it et al.}}},\ }\href {\doibase
  https://doi.org/10.1016/j.physletb.2022.137506} {\bibfield  {journal}
  {\bibinfo  {journal} {Physics Letters B}\ }\textbf {\bibinfo {volume}
  {835}},\ \bibinfo {pages} {137506} (\bibinfo {year} {2022})}\BibitemShut
  {NoStop}%
\bibitem [{\citenamefont {Mayer}(1948)}]{Mayer:1948zz}%
  \BibitemOpen
  \bibfield  {author} {\bibinfo {author} {\bibfnamefont {M.~G.}\ \bibnamefont
  {Mayer}},\ }\href {\doibase 10.1103/PhysRev.74.235} {\bibfield  {journal}
  {\bibinfo  {journal} {Phys. Rev.}\ }\textbf {\bibinfo {volume} {74}},\
  \bibinfo {pages} {235} (\bibinfo {year} {1948})}\BibitemShut {NoStop}%
\bibitem [{\citenamefont {Mayer}(1949)}]{Mayer:1949pd}%
  \BibitemOpen
  \bibfield  {author} {\bibinfo {author} {\bibfnamefont {M.~G.}\ \bibnamefont
  {Mayer}},\ }\href {\doibase 10.1103/PhysRev.75.1969} {\bibfield  {journal}
  {\bibinfo  {journal} {Phys. Rev.}\ }\textbf {\bibinfo {volume} {75}},\
  \bibinfo {pages} {1969} (\bibinfo {year} {1949})}\BibitemShut {NoStop}%
\bibitem [{\citenamefont {Haxel}\ \emph {et~al.}(1949)\citenamefont {Haxel},
  \citenamefont {Jensen},\ and\ \citenamefont {Suess}}]{Haxel:1949fjd}%
  \BibitemOpen
  \bibfield  {author} {\bibinfo {author} {\bibfnamefont {O.}~\bibnamefont
  {Haxel}}, \bibinfo {author} {\bibfnamefont {J.~H.~D.}\ \bibnamefont
  {Jensen}}, \ and\ \bibinfo {author} {\bibfnamefont {H.~E.}\ \bibnamefont
  {Suess}},\ }\href {\doibase 10.1103/PhysRev.75.1766.2} {\bibfield  {journal}
  {\bibinfo  {journal} {Phys. Rev.}\ }\textbf {\bibinfo {volume} {75}},\
  \bibinfo {pages} {1766} (\bibinfo {year} {1949})}\BibitemShut {NoStop}%
\bibitem [{\citenamefont {Lesinski}\ \emph {et~al.}(2007)\citenamefont
  {Lesinski}, \citenamefont {Bender}, \citenamefont {Bennaceur}, \citenamefont
  {Duguet},\ and\ \citenamefont {Meyer}}]{Lesinski:2007ys}%
  \BibitemOpen
  \bibfield  {author} {\bibinfo {author} {\bibfnamefont {T.}~\bibnamefont
  {Lesinski}}, \bibinfo {author} {\bibfnamefont {M.}~\bibnamefont {Bender}},
  \bibinfo {author} {\bibfnamefont {K.}~\bibnamefont {Bennaceur}}, \bibinfo
  {author} {\bibfnamefont {T.}~\bibnamefont {Duguet}}, \ and\ \bibinfo {author}
  {\bibfnamefont {J.}~\bibnamefont {Meyer}},\ }\href {\doibase
  10.1103/PhysRevC.76.014312} {\bibfield  {journal} {\bibinfo  {journal} {Phys.
  Rev. C}\ }\textbf {\bibinfo {volume} {76}},\ \bibinfo {pages} {014312}
  (\bibinfo {year} {2007})},\ \Eprint {http://arxiv.org/abs/0704.0731}
  {arXiv:0704.0731 [nucl-th]} \BibitemShut {NoStop}%
\bibitem [{\citenamefont {Zalewski}\ \emph {et~al.}(2008)\citenamefont
  {Zalewski}, \citenamefont {Dobaczewski}, \citenamefont {Satula},\ and\
  \citenamefont {Werner}}]{Zalewski:2008is}%
  \BibitemOpen
  \bibfield  {author} {\bibinfo {author} {\bibfnamefont {M.}~\bibnamefont
  {Zalewski}}, \bibinfo {author} {\bibfnamefont {J.}~\bibnamefont
  {Dobaczewski}}, \bibinfo {author} {\bibfnamefont {W.}~\bibnamefont {Satula}},
  \ and\ \bibinfo {author} {\bibfnamefont {T.~R.}\ \bibnamefont {Werner}},\
  }\href {\doibase 10.1103/PhysRevC.77.024316} {\bibfield  {journal} {\bibinfo
  {journal} {Phys. Rev. C}\ }\textbf {\bibinfo {volume} {77}},\ \bibinfo
  {pages} {024316} (\bibinfo {year} {2008})},\ \Eprint
  {http://arxiv.org/abs/0801.0924} {arXiv:0801.0924 [nucl-th]} \BibitemShut
  {NoStop}%
\bibitem [{\citenamefont {Bender}\ \emph {et~al.}(2009)\citenamefont {Bender},
  \citenamefont {Bennaceur}, \citenamefont {Duguet}, \citenamefont {Heenen},
  \citenamefont {Lesinski},\ and\ \citenamefont {Meyer}}]{Bender:2009ty}%
  \BibitemOpen
  \bibfield  {author} {\bibinfo {author} {\bibfnamefont {M.}~\bibnamefont
  {Bender}}, \bibinfo {author} {\bibfnamefont {K.}~\bibnamefont {Bennaceur}},
  \bibinfo {author} {\bibfnamefont {T.}~\bibnamefont {Duguet}}, \bibinfo
  {author} {\bibfnamefont {P.~H.}\ \bibnamefont {Heenen}}, \bibinfo {author}
  {\bibfnamefont {T.}~\bibnamefont {Lesinski}}, \ and\ \bibinfo {author}
  {\bibfnamefont {J.}~\bibnamefont {Meyer}},\ }\href {\doibase
  10.1103/PhysRevC.80.064302} {\bibfield  {journal} {\bibinfo  {journal} {Phys.
  Rev. C}\ }\textbf {\bibinfo {volume} {80}},\ \bibinfo {pages} {064302}
  (\bibinfo {year} {2009})},\ \Eprint {http://arxiv.org/abs/0909.3782}
  {arXiv:0909.3782 [nucl-th]} \BibitemShut {NoStop}%
\bibitem [{\citenamefont {Sorlin}\ and\ \citenamefont
  {Porquet}(2012)}]{Sorlin_2013}%
  \BibitemOpen
  \bibfield  {author} {\bibinfo {author} {\bibfnamefont {O.}~\bibnamefont
  {Sorlin}}\ and\ \bibinfo {author} {\bibfnamefont {M.-G.}\ \bibnamefont
  {Porquet}},\ }\href {\doibase 10.1088/0031-8949/2013/T152/014003} {\bibfield
  {journal} {\bibinfo  {journal} {Physica Scripta}\ }\textbf {\bibinfo {volume}
  {2013}},\ \bibinfo {pages} {014003} (\bibinfo {year} {2012})}\BibitemShut
  {NoStop}%
\bibitem [{\citenamefont {Nakada}\ and\ \citenamefont
  {Inakura}(2015)}]{PhysRevC.91.021302}%
  \BibitemOpen
  \bibfield  {author} {\bibinfo {author} {\bibfnamefont {H.}~\bibnamefont
  {Nakada}}\ and\ \bibinfo {author} {\bibfnamefont {T.}~\bibnamefont
  {Inakura}},\ }\href {\doibase 10.1103/PhysRevC.91.021302} {\bibfield
  {journal} {\bibinfo  {journal} {Phys. Rev. C}\ }\textbf {\bibinfo {volume}
  {91}},\ \bibinfo {pages} {021302} (\bibinfo {year} {2015})}\BibitemShut
  {NoStop}%
\bibitem [{\citenamefont {Sharma}\ \emph {et~al.}(1995)\citenamefont {Sharma},
  \citenamefont {Lalazissis}, \citenamefont {Konig},\ and\ \citenamefont
  {Ring}}]{Sharma:1994mim}%
  \BibitemOpen
  \bibfield  {author} {\bibinfo {author} {\bibfnamefont {M.~M.}\ \bibnamefont
  {Sharma}}, \bibinfo {author} {\bibfnamefont {G.}~\bibnamefont {Lalazissis}},
  \bibinfo {author} {\bibfnamefont {J.}~\bibnamefont {Konig}}, \ and\ \bibinfo
  {author} {\bibfnamefont {P.}~\bibnamefont {Ring}},\ }\href {\doibase
  10.1103/PhysRevLett.74.3744} {\bibfield  {journal} {\bibinfo  {journal}
  {Phys. Rev. Lett.}\ }\textbf {\bibinfo {volume} {74}},\ \bibinfo {pages}
  {3744} (\bibinfo {year} {1995})},\ \Eprint
  {http://arxiv.org/abs/nucl-th/9502006} {arXiv:nucl-th/9502006} \BibitemShut
  {NoStop}%
\bibitem [{\citenamefont {Reinhard}\ and\ \citenamefont
  {Flocard}(1995)}]{Reinhard:1995zz}%
  \BibitemOpen
  \bibfield  {author} {\bibinfo {author} {\bibfnamefont {P.~G.}\ \bibnamefont
  {Reinhard}}\ and\ \bibinfo {author} {\bibfnamefont {H.}~\bibnamefont
  {Flocard}},\ }\href {\doibase 10.1016/0375-9474(94)00770-N} {\bibfield
  {journal} {\bibinfo  {journal} {Nucl. Phys. A}\ }\textbf {\bibinfo {volume}
  {584}},\ \bibinfo {pages} {467} (\bibinfo {year} {1995})}\BibitemShut
  {NoStop}%
\bibitem [{\citenamefont {Yue}\ \emph {et~al.}(2024)\citenamefont {Yue},
  \citenamefont {Zhang},\ and\ \citenamefont {Chen}}]{Yue:2024srj}%
  \BibitemOpen
  \bibfield  {author} {\bibinfo {author} {\bibfnamefont {T.-G.}\ \bibnamefont
  {Yue}}, \bibinfo {author} {\bibfnamefont {Z.}~\bibnamefont {Zhang}}, \ and\
  \bibinfo {author} {\bibfnamefont {L.-W.}\ \bibnamefont {Chen}},\ }\href@noop
  {} {\  (\bibinfo {year} {2024})},\ \Eprint {http://arxiv.org/abs/2406.03844}
  {arXiv:2406.03844 [nucl-th]} \BibitemShut {NoStop}%
\bibitem [{\citenamefont {Zhao}\ \emph {et~al.}(2024)\citenamefont {Zhao},
  \citenamefont {Lin}, \citenamefont {Kumar}, \citenamefont {Steiner},\ and\
  \citenamefont {Prakash}}]{Zhao:2024gjz}%
  \BibitemOpen
  \bibfield  {author} {\bibinfo {author} {\bibfnamefont {T.}~\bibnamefont
  {Zhao}}, \bibinfo {author} {\bibfnamefont {Z.}~\bibnamefont {Lin}}, \bibinfo
  {author} {\bibfnamefont {B.}~\bibnamefont {Kumar}}, \bibinfo {author}
  {\bibfnamefont {A.~W.}\ \bibnamefont {Steiner}}, \ and\ \bibinfo {author}
  {\bibfnamefont {M.}~\bibnamefont {Prakash}},\ }\href@noop {} {\  (\bibinfo
  {year} {2024})},\ \Eprint {http://arxiv.org/abs/2406.05267} {arXiv:2406.05267
  [nucl-th]} \BibitemShut {NoStop}%
\bibitem [{\citenamefont {Xu}\ \emph {et~al.}(2015)\citenamefont {Xu},
  \citenamefont {Li}, \citenamefont {Shen},\ and\ \citenamefont
  {Xia}}]{Xu:2015kxa}%
  \BibitemOpen
  \bibfield  {author} {\bibinfo {author} {\bibfnamefont {J.}~\bibnamefont
  {Xu}}, \bibinfo {author} {\bibfnamefont {B.~A.}\ \bibnamefont {Li}}, \bibinfo
  {author} {\bibfnamefont {W.~Q.}\ \bibnamefont {Shen}}, \ and\ \bibinfo
  {author} {\bibfnamefont {Y.}~\bibnamefont {Xia}},\ }\href {\doibase
  10.1007/s11467-015-0479-8} {\bibfield  {journal} {\bibinfo  {journal} {Front.
  Phys. (Beijing)}\ }\textbf {\bibinfo {volume} {10}},\ \bibinfo {pages}
  {102501} (\bibinfo {year} {2015})}\BibitemShut {NoStop}%
\bibitem [{\citenamefont {Xu}\ and\ \citenamefont {Li}(2013)}]{Xu:2012hh}%
  \BibitemOpen
  \bibfield  {author} {\bibinfo {author} {\bibfnamefont {J.}~\bibnamefont
  {Xu}}\ and\ \bibinfo {author} {\bibfnamefont {B.~A.}\ \bibnamefont {Li}},\
  }\href {\doibase 10.1016/j.physletb.2013.06.033} {\bibfield  {journal}
  {\bibinfo  {journal} {Phys. Lett. B}\ }\textbf {\bibinfo {volume} {724}},\
  \bibinfo {pages} {346} (\bibinfo {year} {2013})}\BibitemShut {NoStop}%
\bibitem [{\citenamefont {Xia}\ \emph {et~al.}(2014)\citenamefont {Xia},
  \citenamefont {Xu}, \citenamefont {Li},\ and\ \citenamefont
  {Shen}}]{Xia:2014qva}%
  \BibitemOpen
  \bibfield  {author} {\bibinfo {author} {\bibfnamefont {Y.}~\bibnamefont
  {Xia}}, \bibinfo {author} {\bibfnamefont {J.}~\bibnamefont {Xu}}, \bibinfo
  {author} {\bibfnamefont {B.~A.}\ \bibnamefont {Li}}, \ and\ \bibinfo {author}
  {\bibfnamefont {W.~Q.}\ \bibnamefont {Shen}},\ }\href {\doibase
  10.1103/PhysRevC.89.064606} {\bibfield  {journal} {\bibinfo  {journal} {Phys.
  Rev. C}\ }\textbf {\bibinfo {volume} {89}},\ \bibinfo {pages} {064606}
  (\bibinfo {year} {2014})}\BibitemShut {NoStop}%
\bibitem [{\citenamefont {Xia}\ and\ \citenamefont {Xu}(2020)}]{Xia:2019whr}%
  \BibitemOpen
  \bibfield  {author} {\bibinfo {author} {\bibfnamefont {Y.}~\bibnamefont
  {Xia}}\ and\ \bibinfo {author} {\bibfnamefont {J.}~\bibnamefont {Xu}},\
  }\href {\doibase 10.1016/j.physletb.2019.135130} {\bibfield  {journal}
  {\bibinfo  {journal} {Phys. Lett. B}\ }\textbf {\bibinfo {volume} {800}},\
  \bibinfo {pages} {135130} (\bibinfo {year} {2020})}\BibitemShut {NoStop}%
\bibitem [{\citenamefont {Ring}\ and\ \citenamefont {Schuck}(1980)}]{Ring1980}%
  \BibitemOpen
  \bibfield  {author} {\bibinfo {author} {\bibfnamefont {P.}~\bibnamefont
  {Ring}}\ and\ \bibinfo {author} {\bibfnamefont {P.}~\bibnamefont {Schuck}},\
  }\href@noop {} {\emph {\bibinfo {title} {The Nuclear Many-Body Problem}}}\
  (\bibinfo  {publisher} {Springer},\ \bibinfo {address} {Berlin},\ \bibinfo
  {year} {1980})\BibitemShut {NoStop}%
\bibitem [{\citenamefont {Smith}\ and\ \citenamefont
  {Jensen}(1989)}]{Smith1989}%
  \BibitemOpen
  \bibfield  {author} {\bibinfo {author} {\bibfnamefont {H.}~\bibnamefont
  {Smith}}\ and\ \bibinfo {author} {\bibfnamefont {H.}~\bibnamefont {Jensen}},\
  }\href@noop {} {\emph {\bibinfo {title} {Transport Phenomena}}}\ (\bibinfo
  {publisher} {Oxford University Press},\ \bibinfo {address} {Oxford},\
  \bibinfo {year} {1989})\BibitemShut {NoStop}%
\bibitem [{\citenamefont {Xia}\ \emph {et~al.}(2016)\citenamefont {Xia},
  \citenamefont {Xu}, \citenamefont {Li},\ and\ \citenamefont
  {Shen}}]{Xia:2016xiw}%
  \BibitemOpen
  \bibfield  {author} {\bibinfo {author} {\bibfnamefont {Y.}~\bibnamefont
  {Xia}}, \bibinfo {author} {\bibfnamefont {J.}~\bibnamefont {Xu}}, \bibinfo
  {author} {\bibfnamefont {B.~A.}\ \bibnamefont {Li}}, \ and\ \bibinfo {author}
  {\bibfnamefont {W.~Q.}\ \bibnamefont {Shen}},\ }\href {\doibase
  10.1016/j.physletb.2016.06.029} {\bibfield  {journal} {\bibinfo  {journal}
  {Phys. Lett. B}\ }\textbf {\bibinfo {volume} {759}},\ \bibinfo {pages} {596}
  (\bibinfo {year} {2016})}\BibitemShut {NoStop}%
\bibitem [{\citenamefont {Vautherin}\ and\ \citenamefont
  {Brink}(1972)}]{Vautherin:1971aw}%
  \BibitemOpen
  \bibfield  {author} {\bibinfo {author} {\bibfnamefont {D.}~\bibnamefont
  {Vautherin}}\ and\ \bibinfo {author} {\bibfnamefont {D.~M.}\ \bibnamefont
  {Brink}},\ }\href {\doibase 10.1103/PhysRevC.5.626} {\bibfield  {journal}
  {\bibinfo  {journal} {Phys. Rev. C}\ }\textbf {\bibinfo {volume} {5}},\
  \bibinfo {pages} {626} (\bibinfo {year} {1972})}\BibitemShut {NoStop}%
\bibitem [{\citenamefont {Engel}\ \emph {et~al.}(1975)\citenamefont {Engel},
  \citenamefont {Brink}, \citenamefont {Goeke}, \citenamefont {Krieger},\ and\
  \citenamefont {Vautherin}}]{Engel:1975zz}%
  \BibitemOpen
  \bibfield  {author} {\bibinfo {author} {\bibfnamefont {Y.~M.}\ \bibnamefont
  {Engel}}, \bibinfo {author} {\bibfnamefont {D.~M.}\ \bibnamefont {Brink}},
  \bibinfo {author} {\bibfnamefont {K.}~\bibnamefont {Goeke}}, \bibinfo
  {author} {\bibfnamefont {S.~J.}\ \bibnamefont {Krieger}}, \ and\ \bibinfo
  {author} {\bibfnamefont {D.}~\bibnamefont {Vautherin}},\ }\href {\doibase
  10.1016/0375-9474(75)90184-0} {\bibfield  {journal} {\bibinfo  {journal}
  {Nucl. Phys. A}\ }\textbf {\bibinfo {volume} {249}},\ \bibinfo {pages} {215}
  (\bibinfo {year} {1975})}\BibitemShut {NoStop}%
\bibitem [{\citenamefont {Lenk}\ and\ \citenamefont
  {Pandharipande}(1989)}]{Lenk:1989zz}%
  \BibitemOpen
  \bibfield  {author} {\bibinfo {author} {\bibfnamefont {R.~J.}\ \bibnamefont
  {Lenk}}\ and\ \bibinfo {author} {\bibfnamefont {V.~R.}\ \bibnamefont
  {Pandharipande}},\ }\href {\doibase 10.1103/PhysRevC.39.2242} {\bibfield
  {journal} {\bibinfo  {journal} {Phys. Rev. C}\ }\textbf {\bibinfo {volume}
  {39}},\ \bibinfo {pages} {2242} (\bibinfo {year} {1989})}\BibitemShut
  {NoStop}%
\bibitem [{\citenamefont {Xia}\ \emph {et~al.}(2017)\citenamefont {Xia},
  \citenamefont {Xu}, \citenamefont {Li},\ and\ \citenamefont
  {Shen}}]{Xia:2017dbx}%
  \BibitemOpen
  \bibfield  {author} {\bibinfo {author} {\bibfnamefont {Y.}~\bibnamefont
  {Xia}}, \bibinfo {author} {\bibfnamefont {J.}~\bibnamefont {Xu}}, \bibinfo
  {author} {\bibfnamefont {B.~A.}\ \bibnamefont {Li}}, \ and\ \bibinfo {author}
  {\bibfnamefont {W.~Q.}\ \bibnamefont {Shen}},\ }\href {\doibase
  10.1103/PhysRevC.96.044618} {\bibfield  {journal} {\bibinfo  {journal} {Phys.
  Rev. C}\ }\textbf {\bibinfo {volume} {96}},\ \bibinfo {pages} {044618}
  (\bibinfo {year} {2017})}\BibitemShut {NoStop}%
\bibitem [{\citenamefont {Arndt}\ \emph {et~al.}(1977)\citenamefont {Arndt},
  \citenamefont {Hackman},\ and\ \citenamefont {Roper}}]{PhysRevC.15.1002}%
  \BibitemOpen
  \bibfield  {author} {\bibinfo {author} {\bibfnamefont {R.~A.}\ \bibnamefont
  {Arndt}}, \bibinfo {author} {\bibfnamefont {R.~H.}\ \bibnamefont {Hackman}},
  \ and\ \bibinfo {author} {\bibfnamefont {L.~D.}\ \bibnamefont {Roper}},\
  }\href {\doibase 10.1103/PhysRevC.15.1002} {\bibfield  {journal} {\bibinfo
  {journal} {Phys. Rev. C}\ }\textbf {\bibinfo {volume} {15}},\ \bibinfo
  {pages} {1002} (\bibinfo {year} {1977})}\BibitemShut {NoStop}%
\bibitem [{\citenamefont {Liu}\ and\ \citenamefont
  {Xu}(2024)}]{PhysRevC.109.014615}%
  \BibitemOpen
  \bibfield  {author} {\bibinfo {author} {\bibfnamefont {R.-J.}\ \bibnamefont
  {Liu}}\ and\ \bibinfo {author} {\bibfnamefont {J.}~\bibnamefont {Xu}},\
  }\href {\doibase 10.1103/PhysRevC.109.014615} {\bibfield  {journal} {\bibinfo
   {journal} {Phys. Rev. C}\ }\textbf {\bibinfo {volume} {109}},\ \bibinfo
  {pages} {014615} (\bibinfo {year} {2024})}\BibitemShut {NoStop}%
\bibitem [{\citenamefont {Wiringa}\ \emph {et~al.}(1995)\citenamefont
  {Wiringa}, \citenamefont {Stoks},\ and\ \citenamefont
  {Schiavilla}}]{Wiringa:1994wb}%
  \BibitemOpen
  \bibfield  {author} {\bibinfo {author} {\bibfnamefont {R.~B.}\ \bibnamefont
  {Wiringa}}, \bibinfo {author} {\bibfnamefont {V.~G.~J.}\ \bibnamefont
  {Stoks}}, \ and\ \bibinfo {author} {\bibfnamefont {R.}~\bibnamefont
  {Schiavilla}},\ }\href {\doibase 10.1103/PhysRevC.51.38} {\bibfield
  {journal} {\bibinfo  {journal} {Phys. Rev. C}\ }\textbf {\bibinfo {volume}
  {51}},\ \bibinfo {pages} {38} (\bibinfo {year} {1995})}\BibitemShut {NoStop}%
\bibitem [{\citenamefont {Becattini}\ \emph {et~al.}(2021)\citenamefont
  {Becattini}, \citenamefont {Buzzegoli},\ and\ \citenamefont
  {Palermo}}]{BECATTINI2021136519}%
  \BibitemOpen
  \bibfield  {author} {\bibinfo {author} {\bibfnamefont {F.}~\bibnamefont
  {Becattini}}, \bibinfo {author} {\bibfnamefont {M.}~\bibnamefont
  {Buzzegoli}}, \ and\ \bibinfo {author} {\bibfnamefont {A.}~\bibnamefont
  {Palermo}},\ }\href {\doibase https://doi.org/10.1016/j.physletb.2021.136519}
  {\bibfield  {journal} {\bibinfo  {journal} {Physics Letters B}\ }\textbf
  {\bibinfo {volume} {820}},\ \bibinfo {pages} {136519} (\bibinfo {year}
  {2021})}\BibitemShut {NoStop}%
\bibitem [{\citenamefont {Becattini}\ and\ \citenamefont
  {Karpenko}(2018)}]{PhysRevLett.120.012302}%
  \BibitemOpen
  \bibfield  {author} {\bibinfo {author} {\bibfnamefont {F.}~\bibnamefont
  {Becattini}}\ and\ \bibinfo {author} {\bibfnamefont {I.}~\bibnamefont
  {Karpenko}},\ }\href {\doibase 10.1103/PhysRevLett.120.012302} {\bibfield
  {journal} {\bibinfo  {journal} {Phys. Rev. Lett.}\ }\textbf {\bibinfo
  {volume} {120}},\ \bibinfo {pages} {012302} (\bibinfo {year}
  {2018})}\BibitemShut {NoStop}%
\bibitem [{\citenamefont {Fu}\ \emph {et~al.}(2021)\citenamefont {Fu},
  \citenamefont {Liu}, \citenamefont {Pang}, \citenamefont {Song},\ and\
  \citenamefont {Yin}}]{PhysRevLett.127.142301}%
  \BibitemOpen
  \bibfield  {author} {\bibinfo {author} {\bibfnamefont {B.}~\bibnamefont
  {Fu}}, \bibinfo {author} {\bibfnamefont {S.~Y.~F.}\ \bibnamefont {Liu}},
  \bibinfo {author} {\bibfnamefont {L.}~\bibnamefont {Pang}}, \bibinfo {author}
  {\bibfnamefont {H.}~\bibnamefont {Song}}, \ and\ \bibinfo {author}
  {\bibfnamefont {Y.}~\bibnamefont {Yin}},\ }\href {\doibase
  10.1103/PhysRevLett.127.142301} {\bibfield  {journal} {\bibinfo  {journal}
  {Phys. Rev. Lett.}\ }\textbf {\bibinfo {volume} {127}},\ \bibinfo {pages}
  {142301} (\bibinfo {year} {2021})}\BibitemShut {NoStop}%
\bibitem [{\citenamefont {Liu}\ \emph {et~al.}(2020)\citenamefont {Liu},
  \citenamefont {Sun},\ and\ \citenamefont {Ko}}]{PhysRevLett.125.062301}%
  \BibitemOpen
  \bibfield  {author} {\bibinfo {author} {\bibfnamefont {S.~Y.~F.}\
  \bibnamefont {Liu}}, \bibinfo {author} {\bibfnamefont {Y.}~\bibnamefont
  {Sun}}, \ and\ \bibinfo {author} {\bibfnamefont {C.~M.}\ \bibnamefont {Ko}},\
  }\href {\doibase 10.1103/PhysRevLett.125.062301} {\bibfield  {journal}
  {\bibinfo  {journal} {Phys. Rev. Lett.}\ }\textbf {\bibinfo {volume} {125}},\
  \bibinfo {pages} {062301} (\bibinfo {year} {2020})}\BibitemShut {NoStop}%
\bibitem [{\citenamefont {Wu}\ \emph {et~al.}(2019)\citenamefont {Wu},
  \citenamefont {Pang}, \citenamefont {Huang},\ and\ \citenamefont
  {Wang}}]{PhysRevResearch.1.033058}%
  \BibitemOpen
  \bibfield  {author} {\bibinfo {author} {\bibfnamefont {H.-Z.}\ \bibnamefont
  {Wu}}, \bibinfo {author} {\bibfnamefont {L.-G.}\ \bibnamefont {Pang}},
  \bibinfo {author} {\bibfnamefont {X.-G.}\ \bibnamefont {Huang}}, \ and\
  \bibinfo {author} {\bibfnamefont {Q.}~\bibnamefont {Wang}},\ }\href {\doibase
  10.1103/PhysRevResearch.1.033058} {\bibfield  {journal} {\bibinfo  {journal}
  {Phys. Rev. Res.}\ }\textbf {\bibinfo {volume} {1}},\ \bibinfo {pages}
  {033058} (\bibinfo {year} {2019})}\BibitemShut {NoStop}%
\end{thebibliography}%
\end{document}